\documentclass[traditabstract]{aa} 

\usepackage{graphicx}
\usepackage{txfonts}
\usepackage{color}
\usepackage{natbib}
\usepackage{verbatim}

\newcommand{\cs}{C$_{60}$\,}
\newcommand{\csp}{C$_{60}^+$\,}

\begin{document}

\title{Detection of Buckminsterfullerene emission in the diffuse interstellar medium}
\author{O. Bern\'e \inst{1,2}, N. L. J. Cox \inst{1,2} G. Mulas \inst{3} and C. Joblin \inst{1,2}}
\institute{Universit\'e de Toulouse; UPS-OMP; IRAP;  Toulouse, France
\and CNRS; IRAP; 9 Av. colonel Roche, BP 44346, F-31028 Toulouse cedex 4, France
\and Istituto Nazionale di Astrofisica -- Osservatorio Astronomico di Cagliari -- strada 54, localitˆ Poggio dei Pini, 09012-- Capoterra (CA), Italy}

\titlerunning{Diffuse \cs}
\authorrunning{Bern\'e, Cox, Mulas, Joblin}
\date{Received August ??, 2012; accepted ??, 2012}
\abstract{
{   Emission of fullerenes in their infrared vibrational bands has been detected in space near hot stars. 
The proposed attribution of the diffuse interstellar bands at 9577 and 9632 \AA\ to electronic transitions of the buckminsterfullerene 
cation (i.e. C$_{60}^+$ ) was recently supported by new laboratory data, confirming the presence 
of this species in the diffuse interstellar medium (ISM). In this letter, we present the detection, also in the diffuse ISM, of the 17.4 and 18.9 $\mu$m emission 
bands commonly attributed to vibrational bands of neutral \cs.  According to classical models that compute the charge state of large molecules in space, \cs is expected to be mostly 
neutral in the diffuse ISM. This is in agreement with the abundances of diffuse \cs we derive here from observations. 
}}

 
\maketitle



\section{Introduction} \label{int}

Fullerenes are cage-like macromolecules made of carbon. The most emblematic member of this family, 
\cs, i.e. Buckminsterfullerene, was serendipitously discovered by \citet{kro85} during experiments aimed at simulating
carbon chemistry in the atmospheres of evolved stars. Fullerenes have been extensively studied in the field of chemistry because of their
unique properties and potential applications for nanotechnologies. Soon after the discovery of these macromolecules, it was recognized 
that fullerenes could be present in space and constitute a family of species relevant to astrochemistry \citep{kro85, kro92}. 
{  In particular, it was postulated that their electronic transitions could be at the origin of some of the numerous
unidentified absorption  bands observed towards reddened stars, the so-called  diffuse interstellar bands (DIBs). \citet{foi94}  proposed that
two DIBs at 9577 and 9632 \AA\, could be due to the \cs cation, \csp, but this proposal could not be confirmed at 
that time because of the lack of gas-phase laboratory spectroscopy. 
Clear evidence for the presence of fullerenes (\cs, \csp, C$_{70}$) in space was provided recently with the detection of infrared (IR) emission bands of fullerenes
in regions with intense UV radiation such as evolved stars (e.g. \citealt{cam10, gar10}) and star-forming regions  \citep{sel10, rob12, ber13, cas14}. 
In parallel,  new gas-phase laboratory experiments \citep{cam15} provided an accurate wavelength determination of the electronic
transitions of \csp, which were found to be in excellent agreement with observations, therefore supporting the hypothesis that
the 9577 and 9632 \AA\, DIBs are due to the presence of \csp in the diffuse interstellar medium (ISM). If {  fullerenes} are present in the diffuse 
ISM, they should absorb ultraviolet (UV) photons and re-emit their energy in the IR, hence
their IR vibrational bands should be observed. }
Unfortunately, in the diffuse ISM, the exciting UV radiation field is orders of magnitude smaller compared to star-forming regions, 
resulting in extremely weak IR fluorescence. For this reason fullerene emission has not been detected there so far,
precluding an independent confirmation of their presence in this environment.

\begin{figure*}
\centering
\includegraphics[width=17cm, angle=0]{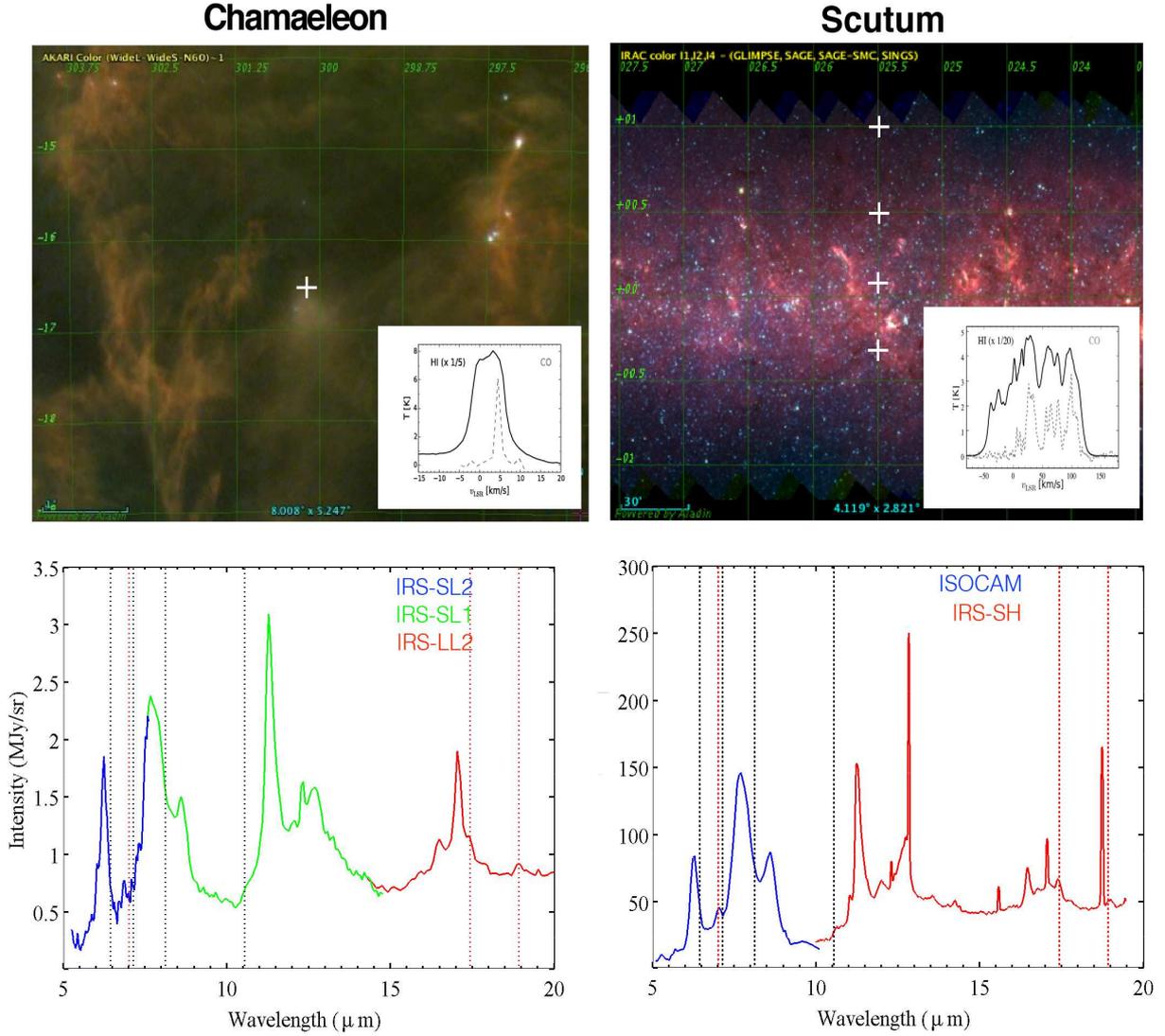}
\caption{  
{  {  Upper panels :} Images of the Chamaeleon (translucent) and Scutum (diffuse) lines of sight studied in this article,
observed with { Spitzer}-IRAC (\citealt{wer04, faz04}). Positions of the spectroscopic observations are 
indicated with white crosses. The images show IRAC colours red=8.0$\mu$m, green=4.5$\mu$m, and blue=3.6$\mu$m. Inserts 
show the HI and CO emission spectra \citep{gas09, dam01, bou98}. 
{  Lower panels}: Mid-infrared spectra towards the studied lines of sight are represented. The vertical lines indicate the positions
of \cs (red) and \csp (black) bands observed in the NGC 7023 reflection nebula \citep{sel10, ber13}. The
band at 7.0 $\mu$m observed in the diffuse line of sight is due to [ArII].}
\label{fig_presentation}}
\label{fig1}
\end{figure*}

\begin{figure*}
\centering
\includegraphics[width=7cm]{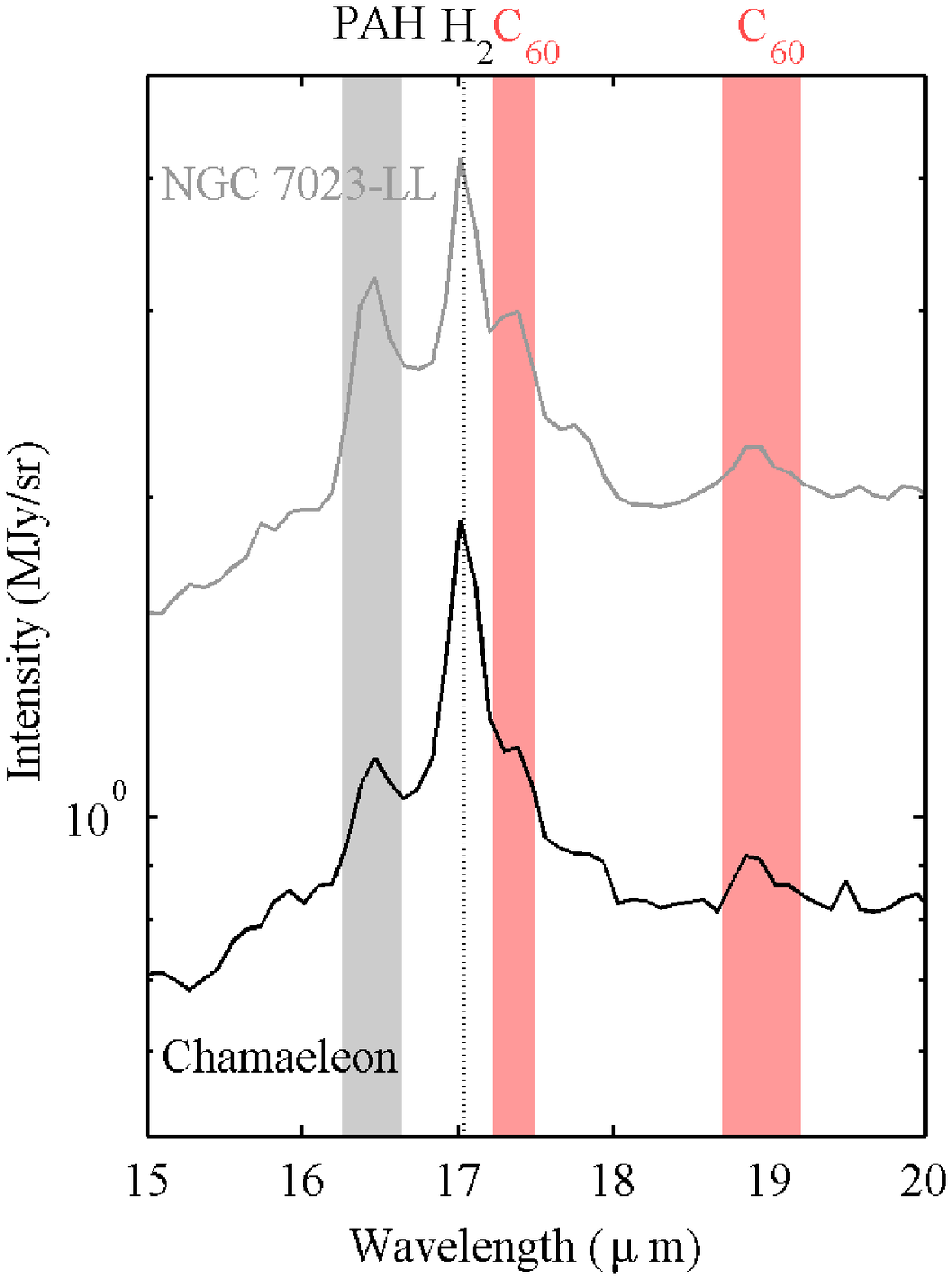}
\includegraphics[width=7cm]{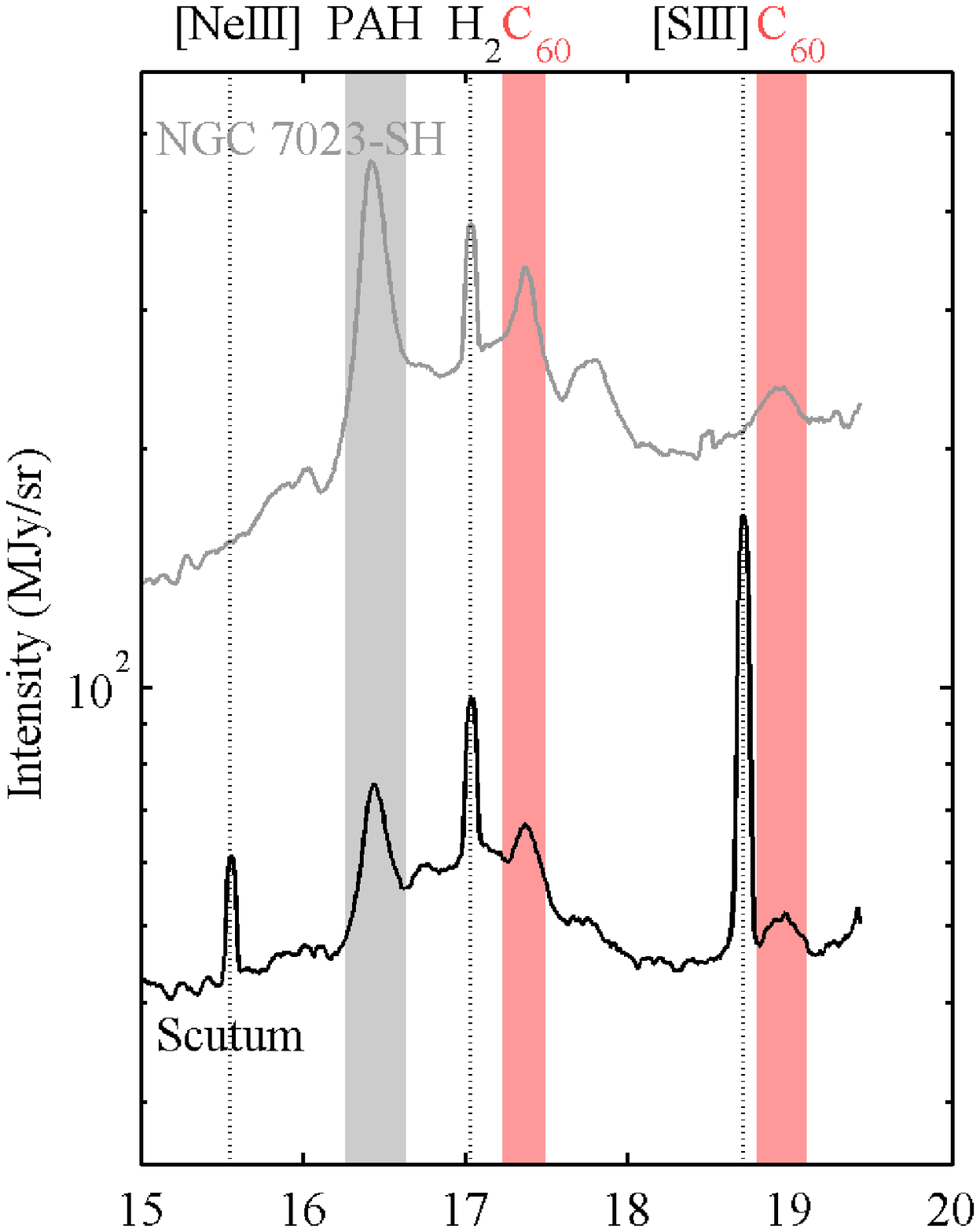}
\caption{ Mid-infrared spectra extracted over the 15-20 $\mu$m range for both lines of sight (see Appendix \ref{app_data} for details). The spectrum of the NGC 7023
nebula where \cs emission has been reported \citep{sel10} is shown as a reference. The Chamaeleon spectrum is compared with 
a low resolution spectrum (IRS-LL) of NGC 7023, while the Scutum line of sight spectrum is compared to a medium resolution spectrum
(IRS-SH) of NGC 7023, to be consistent with the observing modes used to obtain the spectra of interest here (see Appendix \ref{app_data}).
The positions of the \cs vibrational bands are indicated with red vertical patches. The 16.4 $\mu$m attributed to PAH emission is indicated 
with a grey patch. Spectrally unresolved lines of molecular hydrogen and ions are also indicated with vertical dotted lines. \label{fig_spectra}}
\end{figure*}

\begin{table*}
\caption{Properties of the studied lines of sight. \label{tab}}
\begin{center}
\begin{tabular}{lccc}
\hline \hline
                                                                & Chamaeleon                         &  Scutum                       & Refs.  \\
 \hline
 Coord. (Gal. / deg.)                   & +300.13,-16.49         & 26.47 , -0.3--~+1.0        &(1),(2) \\
 UV field  $G_0$$^*$                    &       0.7-3                           &       1-10                            &  (1), (3)  \\
$N_H$ (H~cm$^{-2}$)     & 2.7-5.6$\times10^{21}$ & $5.7\times10^{22}$           &   (1), (3) \\
 A$_V$$^{**}$ (mag)                     & 1.4-3.0                               & 30.5                             &  - \\
 Comment                                                        & Single translucent cloud & Multiple  clouds & - \\
 \hline
\multicolumn{4}{l}{ $^*$\small $G_0=1$ corresponds to a UV field of $1.2\times10^{-7}$ W~m$^{-2}$~sr$^{-1}$.}\\
\multicolumn{4}{l}{ $^{**}$\small  Computed using $N_H/A_V = 1.87 \times 10^{21}$} for Scutum (value for the diffuse ISM) \\
\multicolumn{4}{l}{ \small and  $N_H/A_V = 3.1 \times 10^{21}$ for the Chamaelon (value for translucent/dense ISM).} \\
\multicolumn{4}{l}{ \small  (1) : \citet{ing11}; (2) : this work; (3) : \citet{fal05}} \\

\end{tabular}
\end{center}
\end{table*}

\section{Observations} \label{obs}

{  We  retrieved archival IR spectra obtained with the InfraRed Spectrograph (IRS; \citealt{hou04}) 
on board { Spitzer} \citep{wer04} towards two lines of sight. 
To complement this data, we used the ISOCAM \citep{ces96} data, which was published in
\citet{fla06}. An overview of the studied regions and data are presented in Fig. 1. 

The first line of sight (Fig. 1, left column, { Chamaeleon}), corresponds to a translucent cloud that is situated on the frontside of the Chamaeleon complex at a 
distance of $\sim$70 pc \citep{miz01}. As can be seen from the HI and CO spectra (Fig. 1), this is a single cloud. 
The physical conditions for this cloud have been derived by  \citet{ing11} based on the study of the molecular hydrogen emission:
the hydrogen density is $\sim$ 30-100 particles cm$^{-3}$ and the ultraviolet (UV) radiation field there is 
in the range of $G_0=0.7-3$ (with $G_0=1$ corresponding to $1.2\times10^{-7}$ W~m$^{-2}$~sr$^{-1}$, the 
interstellar standard radiation field calculated by \citealt{hab68}).
The data for this line of sight consists of low resolution spectroscopy ($\lambda/\Delta~\lambda \sim 60-100$) covering 
the 5-35 $\mu$m range and published in \citet{ing11} who focused on H$_2$ emission.
The second line of sight (Fig. 1, right column, { Scutum}) corresponds to a diffuse Galactic region, free of star formation,
situated close to the Scutum.
As can be seen in the HI and CO spectra presented in Fig. 1, this line of sight probes several clouds, where hydrogen can be in molecular, 
atomic, or ionized form.} \citet{fal05} studied this region of the Galaxy in detail using ISO spectroscopy and 
were able to decompose the line of sight into several density components. The radiation field 
on this line of sight varies between $G_0=1$ and $G_0=10$ \citep{fal05}.
The { Spitzer} spectroscopic data for this line of sight was not previously published; it
consists of medium resolution spectroscopy ($\lambda/\Delta~\lambda \sim 600$) covering the 9-20 $\mu$m range.
For both lines of sight, we reduced the data starting from the basic calibrated data and carefully subtracting
background emission (see Appendix \ref{app_data}). The extracted mid-IR spectra over the full spectral range for the two lines of sight are presented in Fig. 1.

In Fig. 2, the same spectra are  shown over a smaller spectral range that is of interest when searching for the emission of \cs. 
For comparison, the spectra of the NGC 7023 massive star-forming region are shown, where \cs was detected by \citet{sel10}
and where the UV radiation field is much higher, i.e. $G_0 = 1000 -10000$. The diffuse and translucent cloud spectra show a number 
of broad features, which are commonly attributed to the emission of large carbonaceous molecules i.e. polycyclic 
aromatic hydrocarbons (PAH; \citealt{tie08}). {  The ratio between the 11.2 and 7.7 $\mu$m PAH bands observed
in both lines of sight is compatible with PAHs being mostly neutral \citep{pil12}.  
The molecular hydrogen lines at 17.0 $\mu$m and 12.3 $\mu$m are also observed towards both lines of sight.} 
In addition, ionic lines from the warm ionized medium are detected towards the Scutum line of sight. 
The two most intense bands of \cs, situated at 17.4 and 18.9 $\mu$m, are 
clearly detected in both lines of sight (Fig. 2).\footnote{The 17.4 $\mu$m can be contaminated by emission from polycyclic aromatic hydrocarbons.}
The vibrational spectrum of \cs is characterized by two additional weaker bands at 7.0 and 8.5 $\mu$m,
which are not detected. However this non-detection is consistent with the 
noise level that exceeds the expected intensity for these bands. The 18.9 $\mu$m band is detected at a 
level that is 3.2 and 12.2 times above the root mean square noise, for the Chameleon 
cloud and for the Scutum line of sight, respectively. 
{  The strongest bands of \csp (at 6.4 and 7.1 $\mu$m, \citealt{ber13}) are not detected. \csp also has an 
emission band close to 18.9 $\mu$m, which could correspond to the emission band detected the spectra in Fig. 2,
but we show in Appendix~\ref{app_csp_em} that \csp cannot contribute more than $\sim$ 25\% to the observed band.}

\section{Abundances of \cs}

Using the integrated intensity of the 18.9 $\mu$m band of \cs and information reported in Table \ref{tab}, 
it is possible to estimate an abundance for \cs in both lines of sight (see Appendix \ref{app_cs_ab}). 
The values found are of the order of
a few $10^{-4}$ to a few 10$^{-3}$ of the gas-phase carbon locked in \cs (Table 2). 
This can be compared to the abundances of \csp for the diffuse ISM derived from DIB measurements.
Using the recent laboratory measurements for the oscillator strength of \csp electronic transitions \citep{cam16},
we derived the abundance of \csp towards lines of sight where the 9577 \AA\, DIB has 
been detected (Appendix \ref{app_csp_ab}). These abundances are in the range of 0.6-1.1$\times10^{-3}$ of the carbon 
(Table~\ref{tb:c60p}), i.e. comparable to what is found for \cs in this study;
they are also in agreement with the upper limit for the abundance of \csp we 
derived from the non-detection of the infrared bands of this species
of 1.8 $\times10^{-3}$ of the carbon (See Appendix \ref{app_csp_ul}). 
{ \bf Abundances for \cs in the diffuse ISM, star-forming regions and evolved stars are summarized in Table 2. 
Fullerenes are most abundant in carbon-rich evolved stars, roughly an order of magnitude less abundant in
the diffuse ISM, and two orders of magnitude less abundant in star forming regions. 
It should however be noted that \cs is detected towards only $\sim$ 3\% of the evolved stars observed with Spitzer 
\citep{ots14}. Overall, the relatively large abundance of \cs in the diffuse ISM may therefore reflect the 
long-term processing of carbonaceous material by UV photons of massive stars, which can lead to the formation of
fullerenes \citep{ber12, ber15}.}


\begin{table}
\caption{Abundances of fullerenes (\% of gas-phase carbon locked in species), derived from emission or absorption measurements in star-forming regions and in the diffuse ISM and evolved stars. \label{ab_ful}}
\begin{center}
\begin{tabular}{lcc}
 \hline
 \hline
                			& Emission                          						& Absorption       \\                                   
\hline
 \multicolumn{3}{c}{\bf Star-forming regions}                                                                   \\
 \hline
 \csp                        	& 0.01$^{*}$            							& -                             \\
 \cs                           	& $0.04-0.06^{**}$     							& -                \\

\hline
				  \multicolumn{3}{c}{\bf Diffuse ISM}                \\
\hline
\csp 				&          0.2 $^{\star}$\textsuperscript{\textdagger}  	& 0.06-0.1 $^{\star}$\\
\cs 				& 0.03-0.4 $^{\star}$                                        		& -                    \\                   
\hline
				  \multicolumn{3}{c}{\bf Evolved stars}     \\
\hline
\csp 				& - 										& 1.2 $^{\star \star }$\\
\cs 				& 0.1-3.0 $^{\star \star \star}$								& - \\
\hline
\multicolumn{3}{l}{ $^*$  \small  From \citet{ber13}; $^{**}$ From \citet{cas14}; } \\
\multicolumn{3}{l}{ $^{\star}$  This work, see Appendices~\ref{app_cs_ab}, \ref{app_csp_ab} and \ref{app_csp_ul};  \textsuperscript{\textdagger} Upper limit;}\\
\multicolumn{3}{l}{ $^{\star \star}$ From \citet{cam11} and references therein} \\
\multicolumn{3}{l}{ $^{\star \star \star}$ From \citet{igl13}, for one source only} \\

\end{tabular}
\end{center}
\end{table}

According to classical theoretical models for the charge 
state of large molecules in space, \cs should be mostly neutral
for standard physical conditions of the diffuse ISM. 
\citet{bak95} computed the charge distribution for the specific case of \cs 
in the diffuse ISM and find that $\sim 30 \%$ is anionic, $\sim 60 \%$ is neutral,
and $\sim 10 \%$ is cationic. These numbers are compatible
with the absolute abundances summarized in Table 2, 
which point to a (not very restrictive) \cs over \csp ratio ranging 
between 0.3 to 6. 
{C$_{60}^-$ is expected to be abundant in models and hence
could be searched for in space. The strongest vibrational bands of C$_{60}^-$ are around  17.5 and 7.3 $\mu$m \citep{kup08}.
The 17.5 $\mu$m band is difficult to identify because of the presence of PAH emission at 
17.4 $\mu$m. A band at  7.3 $\mu$m seems present in the spectrum of the Chamaeleon line of sight, 
however this region of the spectrum is quite noisy and higher sensitivity data will be required to confirm this.}
Observations of diffuse lines of sight with the James Webb Space Telescope, which 
has the sensitivity to detect \cs and \csp as well as tracers of the diffuse ISM should
allow us to conduct more detailed studies on the charge balance of \cs 
in the diffuse ISM. Laboratory studies to quantify key molecular 
parameters involved, such as the electron recombination rate of \csp, are 
also required. 

Finally, the detection of \cs IR emission bands in the diffuse ISM suggests that 
the electronic transitions of this molecules could be identified in absorption. However, given their low oscillator 
strengths, this will be challenging as we illustrate in Appendix \ref{app_cs_dib}.

\section{Conclusion}

In the past few years, our understanding of the organic inventory in space has greatly benefited from
studies of the infrared emission of the Galaxy, in particular with the first identifications of fullerenes. 
Meanwhile, the field of diffuse interstellar bands has been reinvigorated by the convincing assignment of
the 9577 and 9632 \AA\, DIBs to the electronic transitions of \csp. 
The detection of \csp IR emission in the diffuse ISM, with the same abundance as those derived 
from studies in absorption, would provide a strong independent confirmation of 
the DIB assignment. 
It is also likely that astronomical observations combining vibrational and electronic spectroscopy
through the detection of emission and absorption bands, with the support
 of laboratory and theoretical investigations, can help identify new species
 and cast a new light on the DIB conundrum and the organic inventory in space. The detailed spectroscopy of \cs and \csp,
 which are isolated in the diffuse ISM, can also offer the possibility to study some of the fundamental molecular 
 properties of these species in synergy with laboratory and theoretical investigations.

\begin{acknowledgements}
We thank F. Boulanger and E. Falgarone for their thoughtful comments, especially regarding the properties of the studied lines of sight.
We thank Alain Omont, for his detailed comments and for identifying a mistake regarding derivation of abundances.
We thank the anonymous referee whose comments helped improve the manuscript. 
This work was supported by the French programme "Physique et
Chimie du Milieu Interstellaire" (PCMI) funded by the Conseil National de
la Recherche Scientifique (CNRS) and Centre National d'Etudes Spatiales
(CNES). The research leading to these results has also
received funding from the European Research Council under the European Union's Seventh Framework Programme (FP/2007-2013) ERC-2013-SyG, Grant Agreement n. 610256 NANOCOSMOS.
\end{acknowledgements}

\bibliographystyle{aa}
\bibliography{biblio.bib}

{
\appendix

\section{Data analysis}\label{app_data}

The data was retrieved from the { Spitzer} science archive\footnote{http://sha.ipac.caltech.edu/applications/Spitzer/SHA/}.
We used the data at the basic calibrated level (bcd) for both lines of sight. The astronomical observation request
(AOR) keys for these observations are, for the four positions of the Scutum line of sight (LoS): 11060992, 12544768, 11061504,
11061760, from the programme 3513 (PI E. Falgarone), and for the Chamaeleon LoS 28315392, from programme 491 (PI J. Ingalls).
The off positions were included in this latter programme, and for the former observations, we used the offs performed
the same day for extragalactic sources by another programme (programme 1420; PI L. Armus).
The off subtraction, in both cases, was performed at the .bcd level, i.e. before building the 3D spectral cubes.
The data reduction was achieved using the CUBISM software \citep{smi07},
including the slit loss correction function algorithm\footnote{http://tir.astro.utoledo.edu/jdsmith/code/cubism.php}. Once the cubes were 
built, the spectra were obtained by averaging the spectral cubes over all the spatial positions to improve
the signal to noise ratio. 


The complementary data from NGC 7023 has been discussed extensively in the literature, it 
was first presented by \citet{wer04b}. The positions we use to extract the reference spectra
are rectangles whose vertices coordinates are given hereafter:
for the low resolution spectrum (LL) used in the comparison with the translucent cloud spectrum
(21:01:40.579, +68:10:42.96; 21:01:42.343, +68:10:54.6; 21:01:40.255, +68:11:04.44; 21:01:38.491,+68:10:52.80), 
and for the medium resolution spectrum (SH) used in the comparison with the diffuse ISM spectrum
(21:01:24.800, +68:10:11.50; 21:01:32.368, +68:09:33.95; 21:01:40.448, +68:10:24.60; 21:01:32.880,+68:11:02.16).
The LL spectrum was extracted in the north-west photodissociation region (PDR) of the nebula, 
where rotational $H_2$ emission is observed to be strong, i.e. at the surface of the molecular cloud. 
The high resolution spectrum was obtained by averaging the spectral cube over a region that covers the north-west PDR and the cavity of 
atomic gas situated close to the illuminating star HD 200775.

\section{Derivation of neutral \cs abundances from the infrared emission}\label{app_cs_ab}

The derivation of the \cs abundance performed here relies on the assumption that 
the energy absorbed by \cs in the UV is completely re-radiated in the IR bands.
For moderate internal energies ($\sim5-15$ eV), such as those of \cs 
molecules excited by Far-UV photons in the ISM, the main relaxation channel 
is indeed IR emission (see Fig. 2 in \citealt{ber15}) and this assumption is valid.
The total IR intensity in W~m$^{-2}$~sr$^{-1}$ emitted by \cs molecules is hence,
\begin{equation}
I_{tot}=N(C_{60}) \times \sigma_{UV} \times G_0 \times 1.2\times10^{-7},
\label{eqfluo}
\end{equation}
where $N(C_{60})$ is the column density of \cs,  $\sigma_{UV}$ is the UV absorption cross section 
of \cs which we take to be $4.2\times10^{-16}$ cm$^{2}$ following \citet{ber12}, and $G_0$
is the radiation field (see Table 1). Unfortunately, in our observations, the bands at 
shorter wavelengths (7.0 and 8.5 $\mu$m) are not detected, and the 17.4 $\mu$m
band is contaminated by PAH emission, hence $I_{tot}$ cannot be derived directly. 
Instead, it can be estimated  using the 18.9 $\mu$m band intensity ($I_{18.9}$), assuming 
that $I_{tot}=2.3 \times I_{18.9}$, which is what is observed in NGC 7023 \citep{ber13}. 
This assumption is valid considering that the band ratios of the IR bands of \cs depend on the 
average energy of absorbed UV photons that are expected to be comparable
in NGC 7023 and in the diffuse ISM, since both environments
are dominated by the UV fields of young massive stars. 
Hence, one can solve Eq.~\ref{eqfluo} for $N(C_{60})$, and derive the abundance 
of \cs relative to carbon as follows:
\begin{equation}
f_{C}^{C_{60}}=\frac{N(C_{60})\times 60}{N(H)\times[C]},
\label{eqabun}
\end{equation}
adopting a carbon-to-hydrogen fraction measured in the diffuse ISM, i.e. $[C]=1.6 \times 10^{-4}$ \citep{sof04},
and the values for $N(H)$ in Table 1. 
{  For the specific cases considered here, we derive $I_{18.9}=2.5\times10^{-10}$ and $I_{18.9}=1.2\times10^{-8}$ W~m$^{-2}$~sr$^{-1}$
for the Chamaeleon and Scutum LoS respectively. Using Eqs.~\ref{eqfluo}-\ref{eqabun} and the 
values in Table 1 (including uncertainties on the hydrogen column density and radiation field)
yields abundances of \cs of $0.3-2.4 \times 10^{-3}$ and $0.4-3.6  \times 10^{-3}$  of the gas-phase carbon
locked in \cs, for the Chamaeleon and Scutum LoS, respectively. }



\section{C$_{60}^+$ abundances in the diffuse interstellar medium}\label{app_csp_ab}

The fraction of carbon locked in C$_{60}^+$ relative to gas-phase carbon is given by
\begin{equation}
f_C^{C_{60}^+}=\frac{N(C_{60}^+) \times 60}{N(H)\times[C]},
\end{equation}
adopting a carbon-to-hydrogen fraction measured in the diffuse ISM, $[C]=1.6 \times 10^{-4}$.
Values of the column density of \csp, $N(C_{60}^+)$, were derived in previous works along several diffuse 
ISM LoS along which the two DIBs at 9577 and 9632 $\AA$ were observed. In Table C.1, we report the values 
for three well-studied LoS and derived abundances for \csp using the equation above.
The derived values are of the order of a few $10^{-3}$ of the gas-phase carbon. \citet{omo16} find
an average value of the fraction of carbon contained in \csp (with respect to the total carbon abundance, including dust)
 $X(C_{60}^+) = 4\pm2\times10^{-4}$ , equivalent to $f_C^{C_{60}^+}\sim1\times10^{-3}$ of the gas phase carbon, 
 in agreement with our results.
\begin{table*}
\caption{C$_{60}^+$ column densities and abundances (fraction of gas-phase carbon) for several well-studied diffuse ISM lines of sight.}
\label{tb:c60p}
\hspace{1cm}
\begin{tabular}{lllll}
\hline\hline
Line of sight   &       N(C$_{60}^+$)$^*$               & Ref. for 9577 DIB EW & N(H)$^{**}$                        & C$_{60}^+$ abundance $f_C$    \\ \hline
HD169454        &       $0.9 \times 10^{13}$            & \citet{wal15} & $3 \times 10^{21}$              & $1.1 \times 10^{-3}$                  \\ 
HD183143        &       $2 \times 10^{13}$              & \citet{wal15} & $7.4 \times 10^{21}$    & $1.0 \times 10^{-3}$                  \\ 
4U1907+09       &       $2.6 \times 10^{13}$            & \citet{cox14}         & $1.7 \times 10^{22}$    & $6.0 \times 10^{-3}$                  \\
\hline
\multicolumn{5}{l}{ $^*$ \small Column density computed from the measured equivalent width  } \\
\multicolumn{5}{l}{ \small via the measured cross sections given  by \citet{cam16} } \\
\multicolumn{5}{l}{ $^{**}$ \small Hydrogen column density, N(H) = N(H{\sc i} + 2 N(H$_2$), from direct measurements  } \\
\multicolumn{5}{l}{ \small (HD169454: \citealt{dip94}) or visual extinction, $A_V$ (other sightlines; see text for details). } \\

\end{tabular}
\end{table*}

\section{C$_{60}^+$ infrared emission upper limit}\label{app_csp_ul}

It is possible to derive a limit for the abundance of \csp based on the infrared 
spectra of the Chamaeleon and Scutum LoS using the same approach as for \cs (see appendix B). 
{  The 7.1 $\mu$m band of \csp is the easiest to detect \citep{ber13}, however it is absent
in both lines of sight (Fig.~\ref{fig1}). Based on these non-detections, we can place an 
upper limit on $I_{7.1}$, which we can
convert into an $I_{tot}$ for \csp using the 
band ratio observed in NGC 7023 \citep{ber13}.
We derive an upper limit for the \csp column density, which we convert
into an abundance using Eq.~\ref{eqabun}. Using the equation of energy 
budget (Eq.~\ref{eqfluo}) and assuming a similar value of the UV absorption 
cross section for \cs and \csp, we can derive an upper limit for the \csp column 
density, which we convert into an abundance using Eq. C.1.
This yields an upper limit for the \csp abundance 
of $3\times10^{-3}$  of the elemental carbon abundance in the Chamaeleon line of sight
and $\sim 1.8\times 10^{-3}$ of the carbon for the Scutum line of sight. Both limits are above the measurements 
of \csp abundance resulting from the detection of the electronic transitions (Table~\ref{tb:c60p}).}

\section{Prediction of neutral \cs DIB strength}\label{app_cs_dib}

The detection of the IR emission of neutral \cs in the diffuse ISM with abundances comparable to \csp 
suggests that this molecule could be a DIB carrier. Using the 
abundances derived in this paper and relevant data for the
positions, strength, and widths of the electronic transitions of \cs (\citealt{sas01}, Table E1), 
it is possible to compute the predicted depth of the electronic transitions of neutral C$_{60}$. 
Truly accurate positions, oscillator strengths, and band 
widths have not yet been determined for the electronic absorption bands of neutral C$_{60}$ 
in the gas phase at low temperatures, which could therefore be shifted by several \AA\ with 
respect to these recommended values. We use 
the following relation between column density, N, and equivalent width, EW in \AA~\citep{wil37}:
\begin{equation}
\mathrm{EW} = \frac{N\ f\ \lambda^2}{1.13 \times 10^{20}},
\label{eq1}
\end{equation} 
where $\lambda$ is the band position (in \AA) and $f$ its oscillator strength (cf. Table~\ref{tb:sassara}).

For a Gaussian profile the central depth, $\tau_0$, of the line centre relative to the local continuum, is given as 
\begin{equation}
\tau_0 = \frac{EW}{1.571 \times\  \mathrm{FWHM}}
\label{eq2}
.\end{equation}

\begin{table}
\caption{Wavelength, oscillator strength, and FWHM of the considered \cs electronic bands (all values from \citealt{sas01}). }
\centering
\label{tb:sassara}
\begin{tabular}{lll}\hline\hline
$\lambda$ (\AA)         & $f$           & FWHM (\AA)    \\ \hline
3980                    & 0.007 & 5.5                   \\
4024                    & 0.005 & 4.0                   \\
\hline
\end{tabular}
\end{table}

For the diffuse Scutum LoS, the column density derived from infrared emission is
$N$(C$_{60}$) = $5.6 \times 10^{14}$~cm$^{-2}$ for a sightline with total visual extinction, 
$A_V$, of 30.5 magnitudes. For the Chamaeleon LoS, with $A_V$ = 1--2~mag, we have 
$N$(C$_{60}$) = $1.7 \times 10^{13}$~cm$^{-2}$. We can thus infer 
$N$(C$_{60}$) $\sim$ 1$ \times 10^{13}$~cm$^{-2}$/$A_V$ from both sightlines. 
Using Eq.~\ref{eq1} and \ref{eq2}, this yields central depths of $\sim$ 0.1\% per unit visual extinction.
These low values are inherent to the rather low oscillator strengths of the \cs electronic transitions 
and suggest that \cs is unlikely to be responsible for strong DIBs, which can be several percent 
of the continuum \citep{her00}. However, weak DIBs, which can be detected thanks to highly 
sensitive spectroscopy, could well be attributed to \cs with the help of new gas-phase laboratory 
spectroscopy to provide accurate positions, width, and strengths for the electronic transitions.
For instance, on the Scutum LoS, the predicted equivalent width  for the \cs DIBs with
the number quoted above is of the order of 200 provided one can find a proper background star.



\section{Possible contribution of \csp to the 18.9 $\mu$m band}\label{app_csp_em}

{  \csp is known to have an infrared vibrational band at a wavelength close 
to the strongest band of \cs, i.e. at $\sim 18.9$ $\mu$m (see theoretical calculations in 
\citealt{ber13} and experimental data in \citealt{str15}). Based on the upper limits for the 7.1 $\mu$m 
band (Fig. 1), it is possible to derive an estimate of the maximum contribution of \csp to the 
18.9 $\mu$m band (Fig. 2). A proper analysis would require a detailed emission
model, including a description of the UV field, absorption, excitation, and cooling of the molecules.
Here we consider a simplified model. From the observed ratio between the 7.0 and 18.9 $\mu$m 
bands of \cs in NGC 7023 (Sellgren et al. 2010), we can derive a mean colour temperature T$_C$ such that
\begin{equation}
\frac{I_{7.0}}{I_{18.9}}=0.83=\frac{\sigma_{C_{60}}(7.0) \times B(7.0,T_C)}{\sigma_{C_{60}}(18.9) \times B(18.9, T_C)},\end{equation}
where $B(\lambda)$ is the Planck function and $\sigma$ is the infrared cross section 
of the molecule. 
Applying a similar reasoning for \csp and assuming the same colour temperature for both \cs and \csp, the ratio between the 7.1 and 18.9 $\mu$m 
features reads
\begin{equation}
\frac{I_{7.1}}{I_{18.9}}=\frac{\sigma_{C_{60}^+}(7.1) \times B(7.1,T_C)}{\sigma_{C_{60}^+}{18.9} \times B(18.9, T_C)}= \frac{\sigma_{C_{60}^+}(7.1) \times \sigma_{C_{60}}(18.9)}{\sigma_{C_{60}^+}(18.9) \times \sigma_{C_{60}}(7.0) } \times 0.83.
 \label{eqT2}
\end{equation}
Using the IR cross sections for \cs and \csp
of \citet{men00} and \citet{str15}, respectively, yields a theoretical value of $I_{7.1}/I_{18.9}\sim 9$.  
The upper limits for the 7.1 $\mu$m band in the Chamaeleon and Scutum LoS are
$5.9\times10^{-10}$ and $1.2\times10^{-8}$ W~m$^{-2}$~sr$^{-1}$, respectively.
This results in a maximum contribution of \csp to the 18.9 $\mu$m band of
 $6.6\times10^{-11}$ and $1.3\times10^{-9}$ W~m$^{-2}$~sr$^{-1}$, respectively. 
 This is to be compared to the observed emissions of the 18.9 $\mu$m
 band of $2.5\times10^{-10}$ and $1.2\times10^{-8}$ W~m$^{-2}$~sr$^{-1}$.
 Hence, the contribution of \csp to the 18.9 $\mu$m band is at most $\sim$ 25 and 10\%
in the Chamaeleon and Scutum LoS, respectively. In any case, the major contributor to
the 18.9 $\mu$m band is \cs.}

\end{document}